\documentclass[12pt,preprint]{aastex}

\newcommand{\PG} {PG 1115$+$166}

\newcommand{\KUV} {KUV 02196+2816}
\newcommand{\logg} {\log g}
\newcommand{\Te} {T_{\rm eff}}

\newcommand{\msun} {$M_\odot$}
\newcommand\gta{\lower 0.5ex\hbox{$\buildrel > \over \sim\ $}} 
\newcommand\lta{\lower 0.5ex\hbox{$\buildrel < \over \sim\ $}} 
\newcommand{\nh} {N({\rm H})/N({\rm He})}
\newcommand{\nhe} {N({\rm He})/N({\rm H})}
\shortauthors{Limoges et al.}
\shorttitle{Analysis of the White Dwarf \KUV}
\begin{document}

\title{Spectroscopic Analysis of the White Dwarf
\KUV~: \\A New Unresolved DA+DB Degenerate Binary}

\author{M.-M. Limoges, P. Bergeron and Pierre Dufour}
\affil{D\'epartement de Physique, Universit\'e de Montr\'eal, C.P.~6128, 
Succ.~Centre-Ville, 
Montr\'eal, Qu\'ebec H3C 3J7, Canada.}
\email{limoges@astro.umontreal.ca, bergeron@astro.umontreal.ca, dufour@astro.umontreal.ca}

\begin{abstract}

A spectroscopic analysis of the DBA (or DAB) white dwarf \KUV\ is
presented. The observed hydrogen and helium line profiles are shown to
be incompatible with model spectra calculated under the assumption of
a homogeneous hydrogen and helium chemical composition. In contrast,
an excellent fit to the optical spectrum of \KUV\ can be achieved if
the object is interpreted as an unresolved double degenerate composed
of a hydrogen-line DA star and a helium-line DB star. The atmospheric
parameters obtained from the best fit are $\Te=27,170$~K and
$\logg=8.09$ for the DA star, $\Te=36,340$~K and $\logg=8.09$ for the
DB star, which implies that the total mass of the system ($M\sim1.4$
\msun) is very close to the Chandrasekhar limit. Moreover, the effective
temperature of the DB stars lies well within the so-called DB gap
where very few bright DB stars are found. The implications of this
discovery with respect to the DAB and DBA spectral classes and 
to the evolution of double degenerate binaries are discussed.

\end{abstract}


\section{Introduction}

The discovery of double degenerate binary systems is crucial to our
attempt at better constraining their origin and evolution. In
particular, these binary systems are thought to be the progenitors of
Type Ia supernovae if the total mass of the two components exceeds the
Chandrasekhar limit and if their orbital period is small enough to
allow them to merge within a Hubble time. However, only a small
percentage of double degenerate binaries lead to supernova explosions
according to \citet{nel01} and \citet{napi01}. In 2006, only ten
such binaries composed of two white dwarfs were known
\citep{vdS06}. Since then, the SPY (ESO Supernovae Ia Progenitor
surveY) radial velocity survey has observed more than 1000 white
dwarfs and pre-white dwarfs \citep{napi07}. One of the conclusions of
that survey is that double degenerate binaries cannot explain the
existence of every low mass white dwarf (0.45 \msun\ or less). Indeed,
only 42$\%$ of low-mass He-core white dwarfs are found in close binary
systems.

Double degenerate binaries are difficult to detect from single slit
spectroscopy alone. This is particularly true for system composed of
two DA stars since the composite spectrum can be reproduced almost
perfectly with a single hydrogen-rich model atmosphere
\citep{liebert91}. In order to detect such spectroscopically invisible
DA+DA systems, \citet{lajoie07} compared effective temperatures
determined from optical and ultraviolet (IUE) spectra, and identified
three candidate systems. Double degenerate systems composed of white
dwarfs with different spectral types are easier to detect, however,
since the modeling of these objects under the assumption of a single
star leads to spurious results.  This is precisely how \citet{BL02}
showed that the DAB \PG\ was in fact a double degenerate binary system
composed of a DA and a DB star. Their analysis revealed that the
optical spectrum could not be reproduced with a single model
atmosphere with a mixed hydrogen and helium composition, or even a
stratified chemical composition. Instead, the authors showed that a
combination of a DA and a DB model spectrum could perfectly match the
observed line profiles and photometry. Additional objects are also
discussed in \citet{pereira05} and references therein.  \PG\ was first
identified in the PG catalog \citep{green86} as a DA5. Later,
\citet{maxted02} confirmed the binary nature of \PG\ and obtained an 
orbital period of 30.09 days, too long to allow the white dwarfs to
merge within a Hubble time. The binary models of \citet{vdS06}
explained that the older nature of the DB star in \PG\ is the result
of the loss of its hydrogen-rich outer layers after it has evolved
into a giant phase.

We are currently conducting a spectroscopic survey of all the white
dwarfs discovered in the Kiso Ultraviolet (KUV) survey with the aim of
redetermining the luminosity function using the spectroscopic method
of \citet[][see also Liebert et al.~2005]{BSL}. One of these white
dwarfs, \KUV\ (WD~0219$+$282, $V=17.30$) was spectroscopically
identified as a DBA star by
\citet{darling96}. However, this object should have been
classified as a DAB star since the hydrogen lines are actually
stronger than the helium lines (see \S~2 below). DAB white dwarfs are
particularly interesting since they may represent the key to our
understanding of the evolution of DAO and hot DB stars, and of the
existence of a strong deficiency of DB stars between 30,000 and 45,000~K
\citep[see][for a review]{vennes04}.

We report in this paper the discovery that \KUV\ is actually a double
degenerate binary composed of a DA and a DB star. An analysis
identical to that of \PG\ by \citet{BL02} indeed reveals that the
spectrum of \KUV\ cannot be reproduced by a single star model, and
that a composite DA and DB model spectrum provides a much better fit
to the observations. We first present our spectroscopic observations
in \S~2. The model atmosphere calculations described in \S~3 are then
used in \S~4 to analyze in detail the optical spectrum of \KUV\ using
homogeneous as well as composite model atmospheres. Our discussion
follows in \S~5.

\section{Spectroscopic Observations}

\begin{figure}[t]
\begin{center}
\includegraphics[width=40pc]{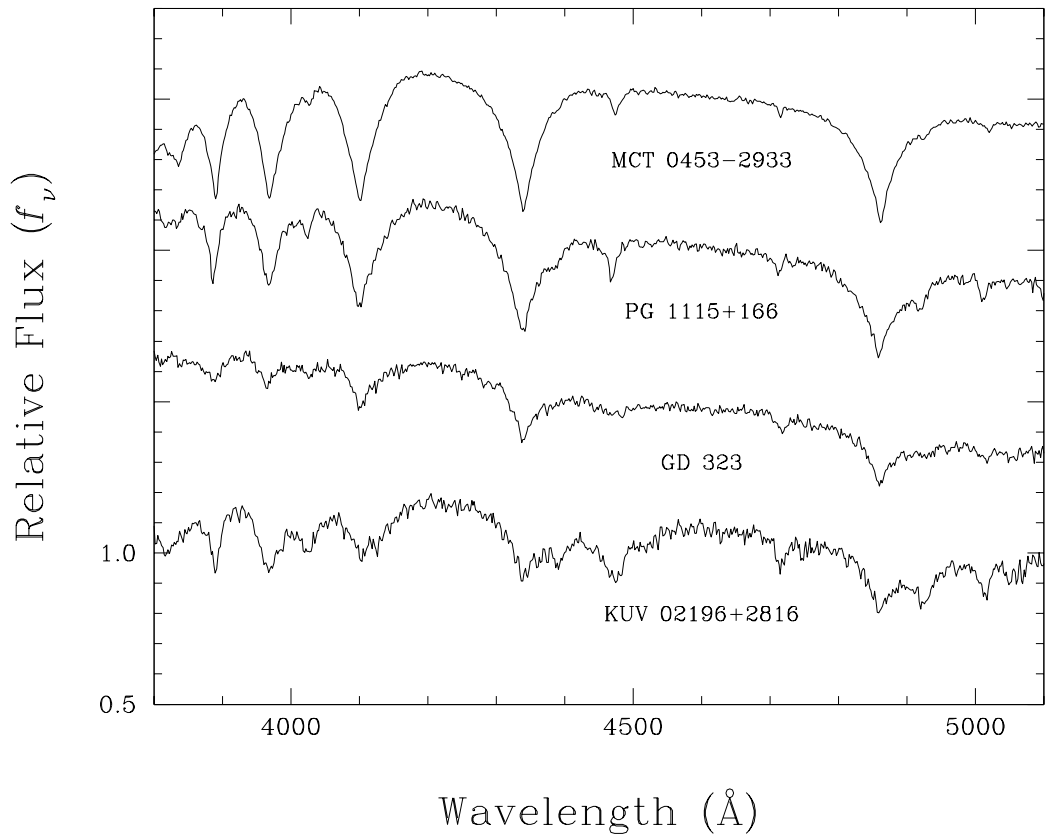}
\end{center}
\caption{Comparison of our optical spectrum of the
DAB star \KUV\ with those of the DAB stars GD 323, \PG, and MCT
0453$-$2933. The last two objects have been interpreted by
\citet{BL02} and \citet{wes94}, respectively,
as unresolved composite systems consisting of a DA white dwarf and a
DB or DBA companion. The spectra are normalized at 4400 \AA\ and are
shifted vertically by 0.5 for clarity.\label{f1}}
\end{figure}

Optical spectroscopy for \KUV\ has been obtained on 2007 December 6
using the Steward Observatory 2.3-m telescope equipped with the Boller
\& Chivens spectrograph and a Loral CCD detector. The
$4.^{\prime\prime}5$ slit together with the 600 line mm$^{-1}$ grating
in first order provided a spectral coverage of 3200--5300 \AA\ at an
intermediate resolution of $\sim 6$~\AA\ FWHM.

Our optical spectrum for \KUV\ is displayed in Figure \ref{f1}. The
hydrogen Balmer lines are clearly stronger than the helium lines and
the star should thus be classified as a DAB white dwarf. The spectrum
of \KUV\ is also compared with GD 323, the prototype of
the DAB class \citep{pereira05}, \PG\ \citep{BL02}, and
MCT 0453$-$2933 \citep{wes94}. \PG\ has already been discussed above
while MCT 0453$-$2933 has been analyzed by \citet{wes94} and was shown
to be an unresolved composite system consisting of a DA white dwarf
together with a DB or a DBA star. From repeated spectroscopic
observations at H$\alpha$ as part of the SPY survey, \citet{napi05}
found variations of the line core shapes.  In one of the spectra, they
noticed a splitting at H$\alpha$
\citep[see also Fig.~7 of][]{voss07}, indicating that both stars in
the binary system contain considerable amounts of hydrogen. It was
concluded that MCT 0453$-$2933 is composed of a DA and a DBA, or even
a DAB star.

The similarity between MCT 0453$-$2933 and \PG\ has already been
discussed in \citet{BL02}, where the double degenerate binary nature
of \PG\ was unveiled. The spectrum of \KUV\ shows the same He~\textsc{i}
features as these two stars: the 4026 and 4471 \AA\ absorption lines
are about the same strength as those of \PG\, while the features at
4713, 4921 and 5015 \AA\ are only slightly stronger. The hydrogen
Balmer lines are definitely weaker in the spectrum of \KUV\ than in
\PG, however, and more typical of the strength observed in GD 323.

\section{Model Atmospheres and Synthetic Spectra}

The model atmospheres and synthetic spectra for DA stars are described
at length in \citet{LBH05} and references therein. The models for DB
and DBA stars used in this analysis rely on a modified version of our
model atmosphere code for DA stars in which we have included all
the helium opacity sources from the DB model atmosphere code described
in \citet{beauchamp96}, and in particular the improved Stark profiles
of neutral helium of \citet{beauchamp97}. The mixed hydrogen and
helium models assume a homogeneous chemical composition. We refrain
here from using stratified models since \citet{BL02} demonstrated that
such models could not improve the fit to the optical spectrum of
\PG\ with respect to homogeneous models. Also, because the hydrogen and helium
lines in the spectrum of \KUV\ are almost of equal strength, DBA
models with unusually high helium abundances had to be
calculated for this analysis. Our DBA model grid covers a range of
$\Te=14,000~(2000)~50,000$~K, $\logg=7.0~(0.5)~8.5$, and
$\log\nhe=-1.0~(0.5)~1.0$, where the numbers in parentheses indicate
the step value. Pure helium models were calculated as well.

\section{Model Atmosphere Analysis}

Our fitting technique relies on the nonlinear least-squares method of
Levenberg-Marquardt \citep{press86}, which is based on a steepest
descent method. The model spectra (convolved with a Gaussian
instrumental profile) and the optical spectrum of \KUV\ are first
normalized to a continuum set to unity. The calculation of $\chi ^2$
is then carried out in terms of these normalized line profiles
only. All atmospheric parameters -- $\Te$, $\logg$ and $\nhe$ -- are
considered free parameters. When fitting DA+DB model spectra, the
total flux of the system is obtained from the sum of the monochromatic
Eddington fluxes of the individual components, weighted by their
respective radius. The stellar radii are obtained from 
evolutionary models similar to those
described in \citet{fon01} but with C/O cores, $q({\rm He})\equiv \log
M_{\rm He}/M_{\star}=10^{-2}$ and $q({\rm H})=10^{-4}$, which are
representative of hydrogen-atmosphere white dwarfs, and $q({\rm He})=10^{-2}$
and $q({\rm H})=10^{-10}$, which are representative of
helium-atmosphere white dwarfs\footnote{see
http://www.astro.umontreal.ca/\~{ }bergeron/CoolingModels/}. 

\begin{figure}[h]
\begin{center}
\includegraphics[width=40pc]{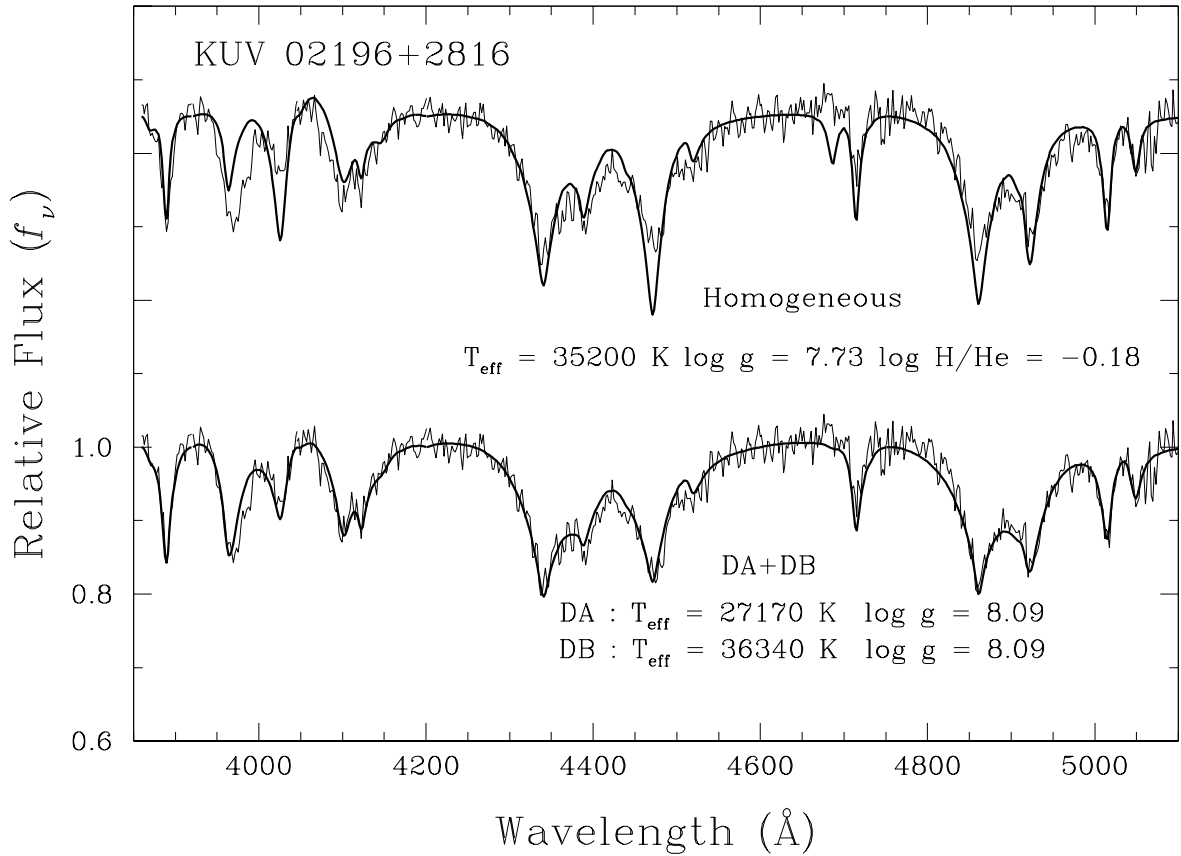}
\end{center}
\caption{Our best fits to the optical spectrum of 
\KUV\ using homogeneous and composite DA+DB
models. The atmospheric parameters for each solution are given in the
figure. Both the observed and theoretical spectra are normalized to a
continuum set to unity. The spectra are shifted by a factor of 0.4
from each other for clarity. Clearly, the DA+DB solution provides the
best fit to the overall spectrum.\label{f2}}
\end{figure}

Our best fits to the optical spectrum of \KUV\ using homogeneous and
composite DA+DB models are shown in Figure \ref{f2}.  The solution
using homogeneous models depicts the same problem as that mentioned in
\citet{burleigh} with the fit to \PG: the He~\textsc{ii} $\lambda$4686 
feature predicted by our models is not observed in the spectrum of
\KUV. We note, however, that \citet{BL02} did not encounter this problem 
while fitting the spectrum of \PG, even though this particular line
was included in their model calculations. Our homogeneous solution for
\KUV\ is $\sim 3000$~K hotter and $\sim0.2$ dex lower in $\logg$ than
that obtained for \PG, but more importantly, the helium abundance is
about a factor of 40 larger, and our solution predicts a weak but
detectable He~\textsc{ii} absorption feature that is simply not
observed. We also see in Figure \ref{f2} that the low hydrogen
Balmer lines (H$\beta$ to H$\delta$) are predicted too strong while
the higher members are too weak. The homogeneous model yields a
satisfactory fit for the He~\textsc{i} lines $\lambda\lambda$4026,
4713, and 5015, but a poor fit to the remaining helium lines. 

\begin{figure}[h]
\begin{center}
\includegraphics[width=40pc]{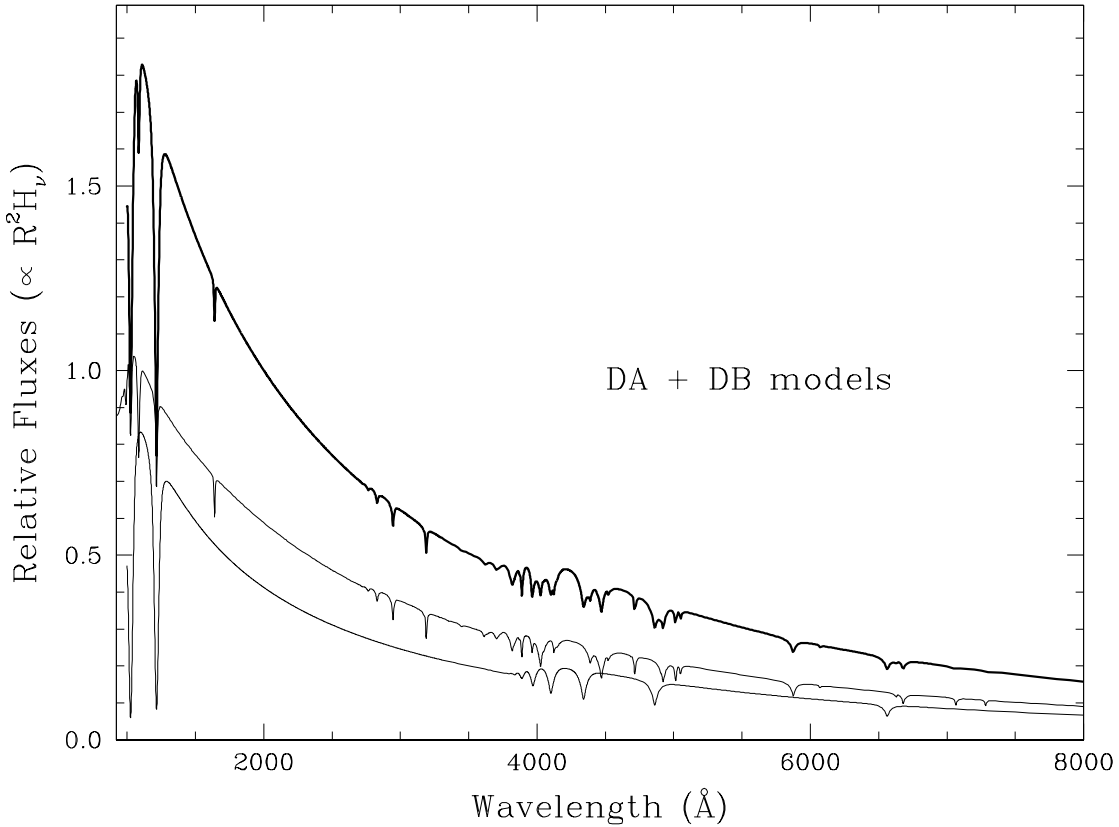}
\end{center}
\caption{Relative energy distributions for our best composite DA+DB fit
displayed in Figure \ref{f2}. The thin lines show the individual
contributions of the DA and DB components, properly weighted by their
respective radius, while the thick line corresponds to the total
monochromatic flux of the composite system.\label{f3}}
\end{figure}

Our grid of model atmospheres with mixed H/He compositions has already
been applied successfully to fit DB and DBA stars
\citep{beauchamp96} or hotter DAO stars \citep{dao94}. Hence the bad fit 
displayed at the top of Figure \ref{f2} does not reflect the inability of our
models to fit this star. We also note that \citet{eisenstein06} have
obtained good fits to DB stars found in the Sloan Digital Sky Survey
(SDSS) with temperatures similar to that of \KUV\ ($\Te=35,200$~K),
although none of these hot DB stars had hydrogen abundances as high as
that inferred here ($\nh\sim 0.7$). Even if we ignore the bad quality
of the fit depicted at the top of Figure \ref{f2}, the atmospheric
parameters we obtain for \KUV\ based on homogeneous models represent a
problem from an astrophysical point of view. Indeed, the typical
hydrogen abundances measured by \citet{voss07} in DBA stars from the
SPY survey are of the order of $\nh\sim10^{-5}$ to $10^{-4}$ (see
their Table 2) --- or even $10^{-3}$ in some extreme cases ---, and
most of them have effective temperatures well below
30,000~K. Similarly, the helium abundances measured in typical DAO and
DAB stars range from $\nhe\sim 10^{-4}$ to 0.1 \citep[see Fig.~7
of][]{vennes04}. It is thus difficult to reconcile the atmospheric
composition determined here for \KUV\ using homogeneous models with
other single DAB/DAO or DBA white dwarfs.

In contrast with the homogeneous solution, our DA+DB solution shown at the bottom
of Figure \ref{f2} provides an excellent fit to the Balmer and neutral
helium lines simultaneously. Our DA+DB fit yields a $\chi^2$ value of
0.543, a value that is significantly lower than that obtained for the
homogeneous solution, $\chi^2=1.416$. All observed features are
reproduced in detail and this is clearly a much better solution for
\KUV. The effective temperature determined for the DA component is
significantly lower than the value achieved under the assumption of a
single object with a homogeneous composition.  This is a direct
consequence of the hydrogen lines being diluted by the continuum flux
of the DB star (see below); the Balmer lines are weakened, and the
effective temperature of the model needs to be increased to match the
observed line profiles. The effective temperature of the DB component,
$\Te=36,340$~K, is higher than the value obtained with our homogeneous
models, $\Te=35,200$~K, yet the He~\textsc{ii} $\lambda$4686 feature
only appears as a small depression in the blue wing of He~\textsc{i}
$\lambda$4713. Again in this case, the presence of the DA star dilutes
all the helium lines present in the DB spectrum. But the main physical
reason for this difference in strength is that the overall opacity of
the mixed H/He model is lower than that of the pure helium model at
the same effective temperature, resulting in lower atmospheric
pressures, which in turn favors the ionization of helium in the mixed
atmosphere. Hence the He~\textsc{ii} $\lambda$4686 feature appears
stronger in the mixed model than in the pure helium model, even though
the former is $\sim 1000$~K cooler.

The surface gravities of both components of the system are almost
identical, $\logg=8.089$ for the DA and $\logg=8.095$ for the DB.
These values can be translated into masses and cooling ages using the
evolutionary models described above. We obtain respectively
for the DA and the DB star 0.69 and 0.68 \msun, and cooling ages of
$1.76\times10^7$ and $5.13\times10^6$ years. Since both stars have the
same radius, the contribution of each component to the total
luminosity is only a function of the effective temperature. Since the
DB star is significantly hotter than the DA component, the former will
contribute more to the combined luminosity of the system. This is
illustrated quantitatively in Figure
\ref{f3} where the contribution of each component to the total
flux is depicted. There is a significant contribution of the flux of
the DB star in the optical regions of the energy distribution. In
particular, the cores of the lower Balmer lines are filled in by the
continuum flux of the DB star, resulting in the poor fits of the
homogeneous solution displayed in Figure \ref{f2} and discussed
above.

\section{Discussion}

Our analysis has shown that the simultaneous presence of hydrogen and
helium in the spectrum of \KUV\ is better explained in terms of an
unresolved binary system composed of a DA white dwarf and a DB star.
We note that in at least two instances, \PG\ and MCT 0453$-$2933
(displayed in Fig.~\ref{f1} and discussed above), this binary
interpretation has been confirmed through spectral velocity
variations, hence similar observations of \KUV\ should also confirm
its binary nature. Despite the evidence of our spectroscopic analysis,
we discuss in the following the difficulty with interpreting \KUV\ as
a single star, even from an evolutionary point of view.

There are actually two scenarios that can produce mixed hydrogen and
helium atmospheres in the temperature range considered here. The first
scenario, discussed in the context of DBA stars, is the accretion of
hydrogen from the interstellar medium onto a helium-dominated
atmosphere. However, the stellar wind model proposed by
\citet{font05} might prevent the accretion of hydrogen for stars
hotter than $\Te\sim20,000$~K. Hence, the presence of hydrogen in
\KUV\ cannot be explained by this scenario, and another explanation
must be sought for the existence of \KUV\ if it is to be interpreted
as a single star.

Another scenario has been proposed by \citet{FW97}, where residual
amounts of hydrogen left in the envelope of hot DO stars would
gradually diffuse to the surface as the white dwarf evolves along the
cooling sequence. This build up of hydrogen at the photosphere would
gradually turn DO white dwarfs into DA stars. Prior to the SDSS, the
absence of any helium-rich atmosphere white dwarfs below
$\Te\sim45,000$~K (and above 30,000~K) would impose a limit of $M_{\rm
H}\ \gta10^{-16}$~\msun\ to the total mass of hydrogen left after the
post-AGB phase. Such minute amounts of hydrogen are actually sufficient
to turn {\it all} hot, helium-rich atmosphere white dwarfs into DA
stars by the time they reach $\Te\sim45,000$~K, which defines the blue
edge of the so-called ``DB gap'', a range in effective
temperature between $\Te\sim30,000$ and 45,000~K where no helium
atmosphere white dwarf (DO or DB stars) had ever been identified
\citep{liebert86}.

More recently, however, the SDSS has revealed the existence of several
hot DB stars in this gap \citep{eisenstein06}, although the fraction
of helium-dominated atmospheres in this temperature range remains
significantly lower than that found at higher or lower temperatures.
We note that SDSS white dwarfs are much fainter than those
investigated by \citet{liebert86} in the Palomar-Green sample.  The
implication of this result is that a few white dwarfs must necessarily
survive the DO to DA transition at high temperatures. This in turn
implies that the amount of hydrogen left in the envelope of pre-white
dwarfs during the post-AGB phase is even smaller than previously
anticipated. We must conclude that the born-again post-AGB scenario
proposed by \citet{WH06} in which a violent mixing event is induced by
a late helium flash in the post-AGB phase is efficient enough to leave
virtually no hydrogen behind.  The recent discovery by
\citet{dufour07,dufour08} of a new class of white dwarf stars with
carbon-rich atmospheres between $\sim18,000$ and 24,000~K even
suggests that these flash events might be efficient enough to leave no
helium behind! In this context, it is perhaps no longer surprising to
find hot DB stars in the gap. These hot DB stars are most likely the
immediate progenitors of the carbon-rich atmosphere white dwarfs
discovered by Dufour et al.

The significant increase in the number of helium-atmosphere DB white
dwarfs below $\Te\sim30,000$~K --- whether in the PG or SDSS samples
--- can only be explained in terms of the convective dilution of the
superficial hydrogen atmosphere by the underlying helium convective
envelope, provided that the hydrogen layer is sufficiently thin
\citep[$M_{\rm H}/M_{\rm tot} \sim 10^{-15}$,][]{macdonald91}.  As
discussed by \citet{BL02}, if the DO-to-DA and DA-to-DB transition
scenarios discussed above are correct, white dwarfs with mixed
hydrogen and helium compositions must be sought either at the hot end
of the DB gap (or deficit) near 45,000~K where an extremely thin
hydrogen atmosphere in diffusive equilibrium on top of the helium
envelope would make the star appear as a DAO star, or near the cool
end of the gap, where convective dilution of the thin hydrogen
atmosphere occurs. Excellent candidates of
such stars caught in the act of changing from one spectral type to
another are PG 1305$-$017, a DAO star at $\Te=44,000$~K whose
spectrum is better reproduced with stratified atmospheres
\citep{dao94}, and GD 323 (shown in Fig.~1), a DAB star at
$\Te=28,750$~K \citep{kls94} whose spectrum exhibits spectroscopic
variations that have been interpreted as surface abundance
inhomogeneities resulting from the convective dilution of hydrogen into
helium \citep{pereira05}.

Going back to \KUV, its effective temperature near 35,000 K obtained
under the assumption of homogeneous single-star models does not fit
well into this picture.  First, the convective efficiency of the
helium envelope at that temperature (the He~\textsc{ii} convection
zone in this case) is much too low to produce any dilution of the
hydrogen superficial atmosphere. And second, since \KUV\ is a full
10,000~K cooler than the blue edge of the DB gap near 45,000~K,
it has necessarily survived the DO-to-DA transition. This in turn
implies that the {\it total} amount of hydrogen left in the envelope
during the post-AGB phase must be extremely small, and certainly too
small to account for the hydrogen abundance inferred from our
homogeneous model atmosphere analysis.

Thus, the DA+DB solution proposed here for \KUV\ not only provides a much
better fit to the optical spectrum than the single star model, but it
also represents the only viable solution from an evolutionary point
of view. Interestingly enough, the effective temperature we infer for
the DB component of the system, $\Te=36,340$~K, puts it right in the
middle of the DB gap. This makes the DB component of the \KUV\ system
the brightest DB white dwarf ever identified in the gap. From the
atmospheric parameters of the DA and DB stars, we estimate the
absolute visual magnitude of the system at $M_V=9.51$, which combined
with the visual magnitude of $V=17.30$ yields a distance of 360 pc.

\citet{wachter03} also found from the 2MASS photometry that
\KUV\ was a good binary candidate composed of a white dwarf and a
low-mass main-sequence star.  This binary nature was confirmed by
\citet{farihi06} in their study of white dwarf-red dwarf systems 
resolved with HST.  Farihi et al.~indeed found that \KUV\ has a red
dwarf companion and both stars are believed to be gravitationally
bound. Since \KUV\ itself is a double degenerate binary, the system is
thus composed of three stars, two of which are unresolved. As
mentioned above, high-resolution spectroscopic observations similar to
that obtained by \citet{napi05} for the double degenerate system MCT
0453$-$2933 could help confirm our binary interpretation. Finally, we
note that since the total mass of the
\KUV\ system ($M\sim1.4$ \msun) is close to the Chandrasekhar limit, a
precise measurement of its orbital period would establish whether
\KUV\ represents a likely supernova candidate.

\acknowledgements We would like to thank the director and staff of Steward Observatory
for the use of their facilities. We would also like to thank
A. Gianninas for a careful reading of this manuscript. This work was
supported in part by the NSERC Canada and by the Fund FCAR
(Qu\'ebec). P.B. is a Cottrell Scholar of Research Corporation for
Science Advancement.

\end{document}